\documentclass[a4paper,12pt]{article}
\usepackage{graphicx}
\begin{document}
\newcommand{\tanb}{\mbox{$\tan \! \beta$}}
\newcommand{\mer}{m_{{\tilde{e}_R}}}
\newcommand{\mlsp}{m_{{\tilde{\chi}^0_1}}}
\newcommand{\sig}{\sigma_{\chi p}}
\newcommand{\mwi}{m_{\tilde{\chi}^+_1}}
\newcommand{\lsp}{\tilde{\chi}^0_1}
\newcommand{\gsim}{\buildrel>\over{_\sim}}
\newcommand{\lsim}{\buildrel<\over{_\sim}}

\setcounter{page}{0}
\thispagestyle{empty}
\begin{flushright}
YITP--00--66 \\
TUM--HEP--394--00 \\
KUNS--1697 \\
November 2000 \\
\end{flushright}

\vspace{2cm}

\begin{center}
{\Large \bf Direct Detection of Neutralino Dark Matter and the
Anomalous Dipole Moment of the Muon}

\baselineskip=32pt

Manuel Drees$^a$, Yeong Gyun Kim$^b$, Tatsuo Kobayashi$^c$ and
Mihoko M. Nojiri$^b$

\baselineskip=22pt

{\it $^a$\, Physik Department, TU M\"unchen, D--85748 Garching, Germany} \\
{\it $^b$\, YITP, Kyoto University, Kyoto, 606-8502, Japan} \\
{\it $^c$\, Physics Department, Kyoto University, Kyoto, 606-8502, Japan}

\end{center}

\vspace{1cm}

\begin{abstract}
\noindent
We compare predictions for the spin--independent contribution to the
neutralino--proton scattering cross section $\sig$ and for the
anomalous magnetic dipole moment of the muon, $a_\mu = (g_\mu-2)/2$,
in models with gravity--mediated supersymmetry breaking. We nearly
always find a positive correlation between these two measurables,
i.e. scenarios with larger $a_\mu$ also tend to have larger $\sig$,
but the detailed prediction differs greatly between models. In
particular, we find that for the popular mSUGRA scenario with
universal soft breaking masses at the scale of Grand Unification,
measurements of $a_\mu$ currently seem more promising. On the other
hand, if scalar soft breaking masses at the GUT scale receive sizable
contributions from $SO(10)$ D--terms, one often finds scenarios with
large $\sig$ but $a_\mu$ below the currently foreseen sensitivity. A
string--inspired model with non--universal scalar spectrum at the GUT
scale falls between these two cases.

\end{abstract}

\vspace{2cm}

\vfill

\pagebreak

\baselineskip=14pt

\section{Introduction}

The Minimal Supersymmetric Standard Model (MSSM) \cite{SUSY} is one of
the best motivated extensions of the Standard Model. It offers a
natural solution of the hierarchy problem \cite{witten} as well as
amazing gauge coupling unification \cite{amaldi}. Since naturalness
requires that at least some superparticles have masses at or below the
TeV scale, supersymmetric theories generally predict a rich
phenomenology at future colliders \cite{colrev}. As extra ``bonus'',
the simplest version of the MSSM, where $R-$parity is conserved, also
contains a new stable particle (the lightest supersymmetric particle,
LSP); in most cases this is the lightest neutralino, which often makes
a good Dark Matter candidate \cite{jungman}.

Unfortunately, direct searches for superparticles at high energy
colliders so far yielded null results. Although it is quite possible
that direct evidence for supersymmetry will be found at the upcoming
``run 2'' of the Tevatron collider, a decisive test of weak scale
supersymmetry may only be possible at the LHC, which will not commence
operations for another five or six years. In the meantime it is
important to explore slightly less direct methods that might either
give evidence for supersymmetry, or else constrain the (usually quite
large) parameter space of supersymmetric models. Here we study two
kinds of experiments which are expected to yield results with greatly
improved sensitivity in the next couple of years: direct searches for
ambient Dark Matter\footnote{Many physicists would accept direct
detection of weakly interacting massive particles (WIMPs) as
(component of) the dark halo of our galaxy as fairly direct evidence
for weak--scale supersymmetry. However, given that direct detection
experiments can basically only measure the mass and scattering cross
section of the WIMP, such a detection, while undoubtedly of the
greatest importance, may not convince everybody that supersymmetry has
been found.}, and the measurement of the anomalous dipole moment of
the muon. 

Current direct Dark Matter search experiments are sensitive to
scenarios with $\tilde{\chi}_1^0 p$ scattering cross section $\sig$ of
order $10^{-6}$ pb or more. In fact, the DAMA experiment even claims
evidence for WIMP Dark Matter, based on the annual modulation of their
counting rate \cite{DAMA}; however, the CDMS experiment, which has
comparable sensitivity, sees no signal \cite{CDMS1}. Within the next
two or three years we can expect results from the upgraded CDMS
experiment (at the deep underground Soudan site) \cite{CDMS2} as well
as from the CRESST experiment in the Gran Sasso laboratory
\cite{cresst}. The sensitivity of these experiments should be about
two orders of magnitude better than that of current experiments, so
that scenarios with $\sig$ exceeding $10^{-8}$ pb might be testable.

SUSY models can also be tested using precision measurements at low
energy experiments. Through their loop effects, sparticles contribute
to low energy physics and these effects may become significant if the
masses of sparticles are not too large. The supersymmetric
contribution to the muon magnetic dipole moment (MDM) $a_\mu$ = ${1
\over 2} (g-2)_\mu$ is one of the most robust probes. Unlike
supersymmetric contributions to rare decays or CP violation in the
$B-$system, which are currently being studied at the $B-$factories
BaBar and BELLE, the supersymmetric contribution to $a_\mu$ is not
very sensitive to the way the supersymmetric flavor and CP problems
are solved.  At present, the muon MDM is measured to be $a_\mu^{exp} =
(116~592~05 \pm 45) \times 10^{-10}~$\cite{carey} and is consistent
with the Standard Model. The ongoing analysis of data from the
Brookhaven experiment E821 is expected to reduce the uncertainty
further and the ultimate goal of the experiment is $\Delta a_\mu \sim
4 \times 10^{-10}~$\cite{e821}.\footnote{In order to fully exploit
such a small experimental error it is also necessary to reduce the
hadronic uncertainty of the SM prediction, which currently stands at
$\sim 7 \times 10^{-10}$ \cite{carey,kino}. Here the analysis of data
that is being collected at the low--energy $e^+e^-$ colliders in
Frascati, Novosibirsk and Beijing, as well as from semi--leptonic
decays of the $\tau$ lepton, will be crucial.} Supersymmetric
contributions can easily exceed this sensitivity without violating any
other constraints on the supersymmetric particle spectrum (from direct
searches or other loop effects) \cite{amugen}.

These supersymmetric contributions to $a_\mu$ stem from
smuon--neutralino and sneutrino--chargino loops. Their size thus
depends on the parameters that appear in the chargino and neutralino
mass matrix, as well as on the masses of the sleptons. Of course, the
parameters of the neutralino mass matrix also determine the mass and
composition of the lightest neutralino, which we assume to form the
Dark Matter in the Universe. Moreover, at least for the theoretically
most appealing case of a bino--like LSP, the LSP relic density is
essentially determined by the mass of the LSP as well as the masses of
the $SU(2)$ singlet sleptons \cite{dn3}. The predictions for the relic
density of bino--like LSPs and for the supersymmetric contribution to
$a_\mu$ are thus related.

There also exist relations between the SUSY contribution to $a_\mu$
and the predicted LSP--nucleon scattering cross section. As already
mentioned, both quantities depend on the parameters of the neutralino
mass matrix. An even more direct connection comes from the common
dependence on the ratio \tanb\ of the vevs of the two neutral Higgs
fields of the MSSM, since both $a_\mu$ and $\sig$ increase with
increasing \tanb. The operator in the effective Lagrangian that gives
rise to the anomalous magnetic moment of the muon couples left-- and
right--handed muons, i.e. violates chirality. In the MSSM (as well as
in the SM) all chirality violation in the (s)muon sector is
proportional to the Yukawa coupling of the muon. This coupling in turn
is proportional to $1/\cos \! \beta$, which scales like \tanb\ if
$\tan^2\beta \gg 1$. The leading contributions to the
spin--independent (coherent) contribution to $\sig$ also involve
violation of chirality \cite{dn5}, but this time in the (s)quark
sector.\footnote{There are also squark exchange contributions where
chirality violation is provided by the LSP mass, but these
contributions are suppressed by an extra power of $(m^2_{\tilde q} -
m^2_{\tilde \chi})^{-1}$ compared to the leading squark exchange
contribution, and can thus usually be ignored in the type of model we
are considering \cite{dn5}.} These contributions either come from the
exchange of CP--even Higgs bosons, or from squark exchange. In both
cases chirality violation ultimately comes from the quark Yukawa
coupling, which, in case of down--type quarks, has the same \tanb\
dependence as the Yukawa coupling of the muon. Note that searches for
Higgs bosons at LEP already imply $\tanb \gsim 2$ in the
MSSM\footnote{Higgs searches also permit $\tanb \lsim 0.6$, but this
is strongly disfavored on theoretical grounds, since the top Yukawa
coupling would have a Landau pole quite close to the weak scale.};
this implies that the Yukawa couplings of up--type quarks, which scale
$\propto 1/\sin\beta$, must be close to their SM--values, and
essentially independent of \tanb. The upshot of this discussion is
that both the SUSY contribution to $a_\mu$ \cite{amugen,moroi} and
$\sig$ \cite{sigold,dn5,signew} can be quite large if $\tanb \gg 1$.

In this paper, we investigate these connections between the SUSY
contribution $\Delta a_\mu$ to the magnetic dipole moment of the muon
and the spin--independent neutralino--proton cross section
quantitatively. We study the minimal supergravity (mSUGRA) model as
well as some other SUSY models where the assumption of strict scalar
mass universality at the GUT scale is relaxed.  In Sec. 2, we discuss
$\Delta a_\mu$ and $\sig$ in the mSUGRA scenario.  These observables
are below the sensitivity of near--future experiments in the small
$\tan\beta$ region, partly because the experimental Higgs mass bounds
then tightly restrict the allowed SUSY parameter space. However, they
can become significant, and could be detectable, if \tanb\ is
large. In Sec. 3 we relax the assumption of universal scalar
masses. We find that the correlation between $\Delta a_\mu$ and $\sig$
quite sensitively depends on the boundary condition at the GUT
scale. Sec. 4 is devoted to conclusions.

\section{mSUGRA}

In the minimal supergravity model, it is usually assumed that all
squared scalar masses receive a common soft SUSY breaking contribution
$m^2_Q = m^2_U = m^2_D = m^2_L = m^2_E = m^2_{H_u} = m^2_{H_d} \equiv
m^2$ at the GUT scale $M_X \simeq 2 \cdot 10^{16}$ GeV, while all
gauginos receive a common mass $M$ and all trilinear soft terms unify
to $A$. The renormalization group (RG) evolution of soft breaking
squared Higgs masses then leads to consistent breaking of the
electroweak symmetry, provided the higgsino mass parameter $\mu$ can
be tuned independently \cite{radbreak}. In this paper, we chose the
weak scale input parameters $m_b(m_b)=4.2$ GeV, $m_t(m_t)=165$ GeV,
and $\tan\beta$. We minimize the tree level potential at
renormalization scale $Q= \sqrt{m_{\tilde{t}}m_t}$, which essentially
reproduces the correct value of $\mu$ obtained by minimizing the full
1--loop effective potential \cite{dn2}.

With these assumptions, the mSUGRA model allows four continuous free
parameters ($m, M, A$ and $\tan\beta$). Also, the sign of $\mu$
remains undetermined. In Fig. 1, we plot the SUSY contribution to the
muon MDM, $\Delta a_\mu$, vs. the spin--independent neutralino-proton
cross section, $\sig$, for three different choices of $\tanb = 4, 10$
and 30. Here, we take $A=0$ and $\mu > 0$ and allow $m$ and $M$ to
vary in the intervals $m < 500$ GeV and $M < 600$ GeV. We include loop
corrections to the masses of neutral Higgs bosons from third
generation quarks and squarks, including leading two--loop corrections
\cite{2loop}. We use the expressions of ref.\cite{moroi} for the
calculation of $\Delta a_\mu$. The calculation of $\sig$ is based on
refs. \cite{dn5,dd}. We use the value $m_s \langle p | \bar{s} s | p
\rangle = 130$~MeV for the strange quark's contribution to the nucleon
mass; this matrix element is uncertain to about a factor of 2, leading
to a similar uncertainty in the prediction of $\sig$. Finally, the
calculation of the scaled LSP relic density $\Omega_\chi h^2$ uses
results of refs.\cite{dn3,bdd}; $s-$channel poles are treated as
described in ref.\cite{dy}. 
The co--annihilation of $\lsp$ with sleptons is not included. This
effect can increase the cosmologically allowed region of parameter
space towards higher LSP masses, if $\mlsp \simeq m_{\tilde l}$;
however, in this (limited) region of parameter space both $\Delta
a_\mu$ and $\sig$ are well below present and near--future sensitivity.
%
\begin{figure}[htbp]
\begin{center}
\includegraphics[width=8.5cm,angle=0]{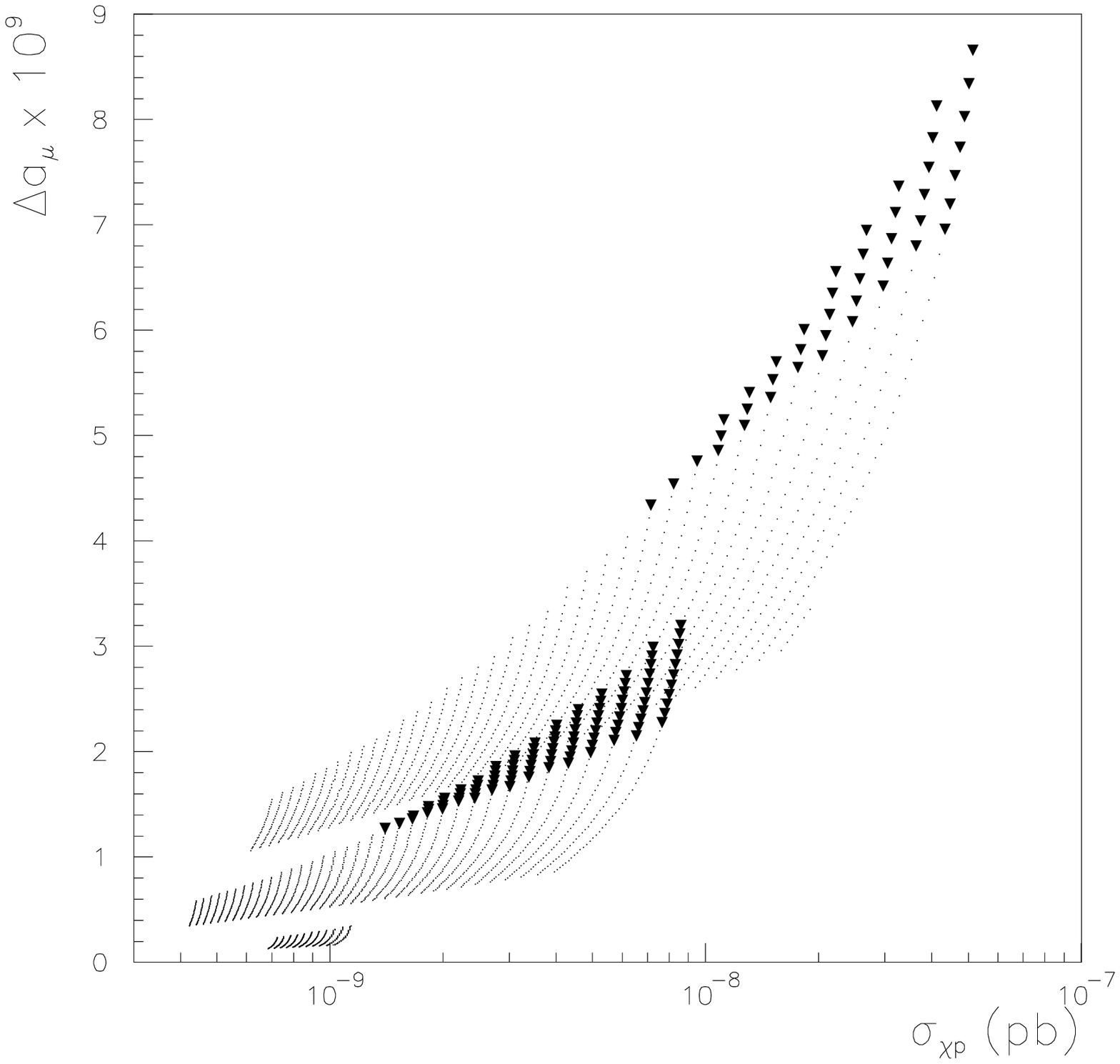}
\end{center}
\caption{\footnotesize $\Delta a_\mu$ vs. $\sig$ in mSUGRA
for $\tanb = 4, 10, 30$ (from bottom to top region). We take $A = 0$
and $\mu > 0$, and scan $m \in [0,500]$ GeV and $M \in [0, 600]$ GeV,
subject to experimental constraints. The heavily marked points satisfy
the requirement $\Omega_\chi h^2 < 0.3$.}
\label{fig1}
\end{figure}

Accelerator bounds significantly limit the SUSY parameter space. For
the parameter range scanned in this plot we find that the lightest
Higgs boson $h$ couples essentially like the single Higgs boson of the
SM; we thus demand that its mass $m_h >$ 111 GeV. This follows from
recent LEP results \cite{lephiggs}, allowing for a 2 GeV theoretical
uncertainty in the calculation of $m_h$ \cite{2loop}. We further
require that the chargino mass $\mwi >$ 100 GeV \cite{alnew}. We
exclude regions where the LSP is charged, i.e.  where $m_{\tilde
\tau_1} < m_{\tilde{\chi}_1^0}$ or $m_{\tilde{t}_1} <
m_{\tilde{\chi}_1^0}$. The heavily marked points satisfy the further
requirement $\Omega_\chi h^2 < 0.3$. This bound on the matter density
in the Universe follows from the analysis of recent cosmological data
\cite{ombound}.

The lowest dotted region in Fig. 1 corresponds to $\tanb=4$. In
this case both $\Delta a_\mu$ and $\sig$ are too small to be
detected in the near future. Furthermore, the lower bound on the Higgs
mass forces one into a region of parameter space where the dark matter
density is quite large and therefore cosmologically disfavored.
Through radiative corrections, which increase with increasing stop
masses, $m_h$ depends quite strongly on $M$, but only weakly on $m$.
The reason is that contributions from Yukawa interactions to the RG
equations reduce scalar masses at the weak scale as compared to their
GUT scale values. As a result, in the expressions for the weak scale
squared stop masses $m^2_{\tilde{t}_L}$ and $m^2_{\tilde{t}_R}$, $m^2$
appears with coefficients significantly less than 1, while $M^2$
appears with coefficients well above 1 \cite{radbreak}. For $\tanb=4$,
the Higgs mass bound $m_h > 111$ GeV requires $M > 500$ GeV, which
corresponds to $\mlsp > 217$ GeV. Such a large SUSY mass scale gives
too large a DM mass density \cite{dn3}.

As $\tan\beta$ increases, both $a_\mu$ and $\sig$ increase. As
mentioned above there are essentially two types of diagrams involving
superparticles which contribute to $a_\mu$, i.e., neutralino--smuon
and chargino--sneutrino loop diagrams.  Since $a_\mu$ requires
chirality violation, for $\tanb \gg 1$ the dominant contributions are
proportional to the product of an electroweak gauge coupling and the
Yukawa coupling of the muon, where the latter factor either comes
directly from the higgsino component of the chargino or neutralino in
the loop, or from $\tilde{\mu}_L - \tilde{\mu}_R$ mixing. For $\tanb
\gg 1$ one thus has $\Delta a_\mu \propto \tanb$.  If mass splittings
between different sparticles are not too large, so that their mass
scale can be described by the single parameter $m_{\rm SUSY}$, the
total result can be estimated as \cite{moroi}
\begin{eqnarray} \label{amu}
\left| \Delta a_\mu \right| = {1 \over 32 \pi^2} \left( \frac{5}{6}
g_2^2 + \frac{1}{6} g_1^2 \right)
{m_\mu^2 \over m_{\rm SUSY}^2} \tan\beta
\end{eqnarray}
where $g_1$ and $g_2$ are the $U(1)_Y$ and $SU(2)$ gauge couplings,
respectively. In our convention, where the gaugino masses and \tanb\
are positive, the sign of $\Delta a_\mu$ is equal to the sign of
$\mu$.

On the other hand, three classes of diagrams contribute to LSP--quark
scattering: the exchange of a $Z$ or CP--even Higgs boson in the $t$
channel and squark exchange in the $s$ or $u$ channel. LSP
interactions with matter can naturally be separated into
spin--dependent and spin--independent parts. $Z$ and squark exchange
contribute to the former, and squark and Higgs exchange to the latter.
In most situations the dominant contribution to the spin independent
amplitude is the exchange of the two neutral CP--even Higgs bosons,
although in some cases the contribution of squark exchange is
substantial \cite{dn5}. For given inputs $m, M, A$ and sign$(\mu)$,
the mass of the heavy CP--even Higgs boson decreases as $\tan\beta$
increases. This effect originates from the contribution of the bottom
Yukawa coupling to the RGE of $m^2_{H_d}$ \cite{dn1}; recall that this
Yukawa coupling also scales like \tanb\ for $\tanb \gg 1$. Moreover,
its coupling to $d-$type quarks is essentially proportional to \tanb\
for $\tan^2 \beta \gg 1$ and $m_A^2 > m^2_{h,{\rm max}}$, where $m_A$
is the mass of the CP--odd Higgs boson and $m^2_{h,{\rm max}}$ is the
$m_A \rightarrow \infty$ limit of the mass of the light CP--even Higgs
boson. Under these conditions, which are almost always satisfied in
the models we are studying, $h$ couples essentially like the single
Higgs boson of the SM, independent of the value of \tanb; $m_h$ also
becomes almost independent of \tanb\ for $\tanb \gsim 10$. The
contribution to $\sig$ from $H$ exchange, which grows quickly with
increasing \tanb, thus starts to dominate over that from $h$ exchange
once $\tanb \gsim (m_H/m_h)^2$. Moreover, the leading down--type
squark exchange contribution to $\sig$, which requires chirality
breaking in the (s)quark sector, also increases $\propto \tanb$ for
large \tanb.

We can see these enhancements in Fig. 1. For $\tanb = 10, \ \Delta
a_\mu$ is, in the cosmologically favored region, well above $1 \times
10^{-9}$ and can reach upto $\sim 3 \times 10^{-9}$, which could be
detected in the near future.  Also, $\sig$ increases quite a lot, but
still remains below $10^{-8}$ pb. These enhancements, compared to the
$\tanb = 4$ case, partly come from the lower SUSY mass scale that is
permitted by the Higgs search limit. For $\tan\beta=10$, the Higgs
mass bound $m_h > 111$GeV only implies $M > 260$ GeV, which
corresponds to $\mlsp > 110$ GeV; this allows scenarios with
cosmologically favored DM mass density in the low $m$ and $M$
region. For $\tan \beta=30$, in the cosmologically favored region
$\Delta a_\mu$ takes values of $ 4 \sim 9 \times 10^{-9}$, while
$\sig$ is mostly above $1 \times 10^{-8}$ pb and can reach upto $5
\times 10^{-8}$ pb. The smallest allowed SUSY mass scale for
$\tanb=30$ is only slightly lower than that for $\tanb=10$, since, as
stated above, $m_h$ becomes quite insensitive to \tanb\ for $\tanb
\gsim 10$.

Note that the highest $\Delta a_\mu$ values shown in Fig.~1 already
reach the {\em current} level of sensitivity, whereas the maximal
value of $\sig$ in this figure is still more than an order of
magnitude below current sensitivity. We conclude that, in view of near
future detectability, $\Delta a_\mu$ is more sensitive than $\sig$ in
the mSUGRA model; this agrees with results of ref.\cite{fenga} (for $m
\leq 500$ GeV). One of the important factors for determining the size
of $\sig$ is the higgsino component of the LSP. MSUGRA predicts a
bino--like LSP $\lsp$ for moderate values of $m$ and $M$ (below $\sim
500$ GeV) \cite{an1,dn3}. This is a rather model independent result
\cite{falk}. Large positive corrections to squark masses from gaugino
loops, together with the large top Yukawa coupling, drive the squared
soft breaking Higgs mass $m^2_{H_u}$ negative at the weak scale. On
the other hand, correct symmetry breaking requires $m^2_{H_u}+\mu^2>
-M_Z^2/2$. One has to make $|\mu|$ large in order to obtain the
correct electroweak symmetry breaking scale, if scalar masses and
gaugino masses are of the same order. Since the Higgs--LSP--LSP
couplings require higgsino--gaugino mixing, they scale like $1/\mu$
for $\mu^2 \gg M_Z^2$. The Higgs exchange contribution to $\sig$ then
scales like $1/\mu^2$. Similarly, the prediction for $\Delta a_\mu$
depends on $\mu$ through gaugino--higgsino mixing as well as
$\tilde{\mu}_L - \tilde{\mu}_R$ mixing. The results of Fig.~1 thus
depend quite sensitively on the value of $\mu$ at the weak scale,
which in turn depends on the GUT scale boundary conditions. The
correlation between $\Delta a_\mu$ and $\sig$ could also be different
in generalizations of mSUGRA. That is the subject of the next section.

Before going to the next section, we wish to discuss the dependence of
our results on the parameter $A$ as well as the sign of $\mu$. The
$A-$dependence is mostly indirect. First of all, $A$ has some impact
on $m_h$, and hence on the allowed region of the ($m, M$) plane. This
can be seen in Fig. 2 a), where the Higgs mass dependence on $M$ is
plotted for three different choices of $A=+2 M, 0, -2 M$. For a given
$M$, a larger $A$ gives larger $m_h$. Therefore, if we chose a large
positive $A$ value, the minimal allowed value of $M$ will be lowered,
so that the maximal allowed value of $\Delta a_\mu$ will be raised,
even though this observable is almost independent of $A$ if the
slepton and --ino mass parameters are held fixed. $A$ also has some
impact on the $\mu$ parameter. As we can see in Fig. 2 b), larger $A$
values give higher $\mu$ values and therefore more bino--like LSP for
a given $M$. This reduction of the higgsino component of the LSP would
reduce the Higgs exchange contribution to $\sig$. Some numerical
examples of the effects of varying $A$ are given in the next section.
\begin{figure}[htbp]
\hskip 0.0cm
\includegraphics[width=8.5cm,angle=0]{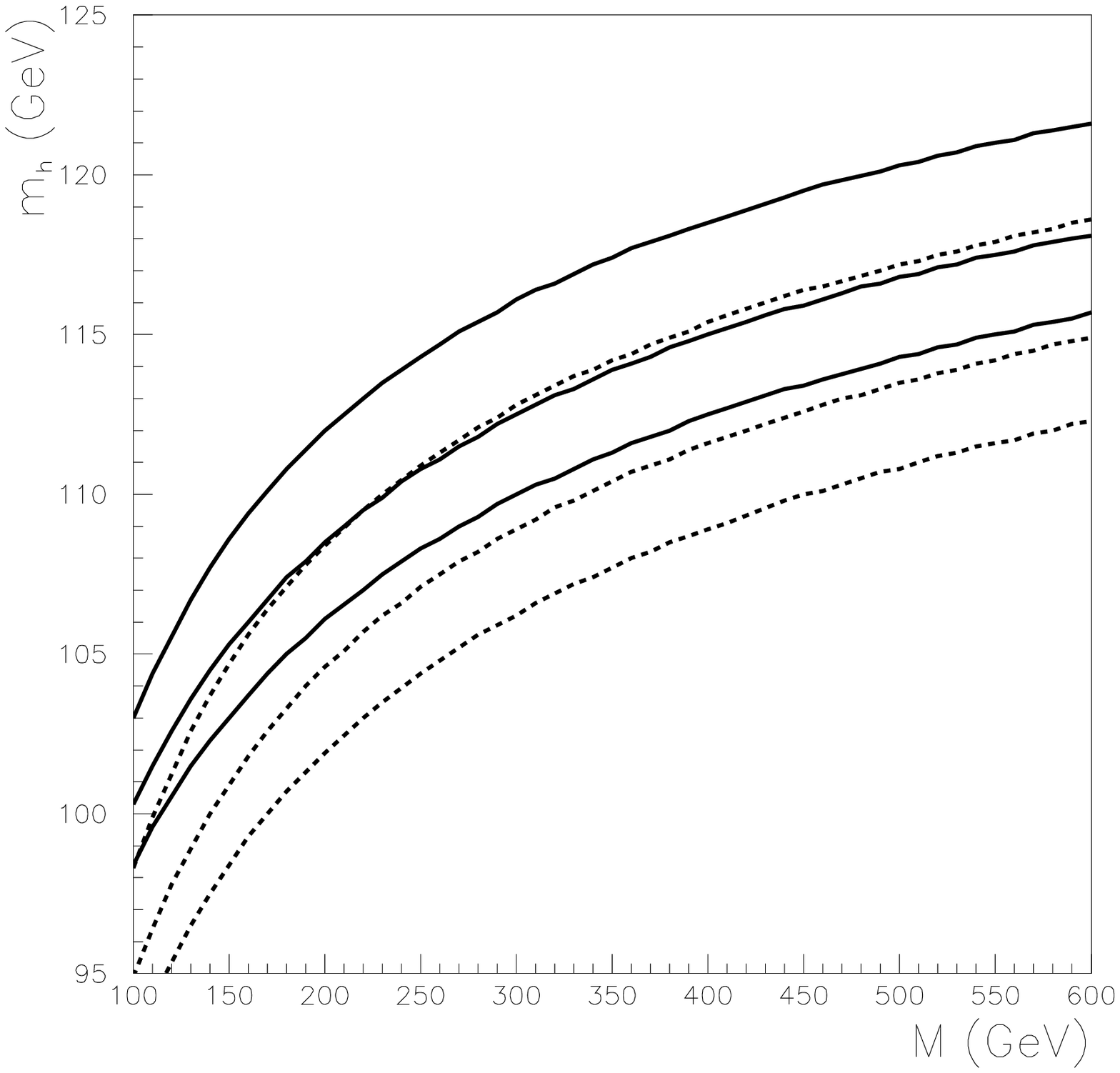}
\hskip 0.5cm
\includegraphics[width=8.5cm,angle=0]{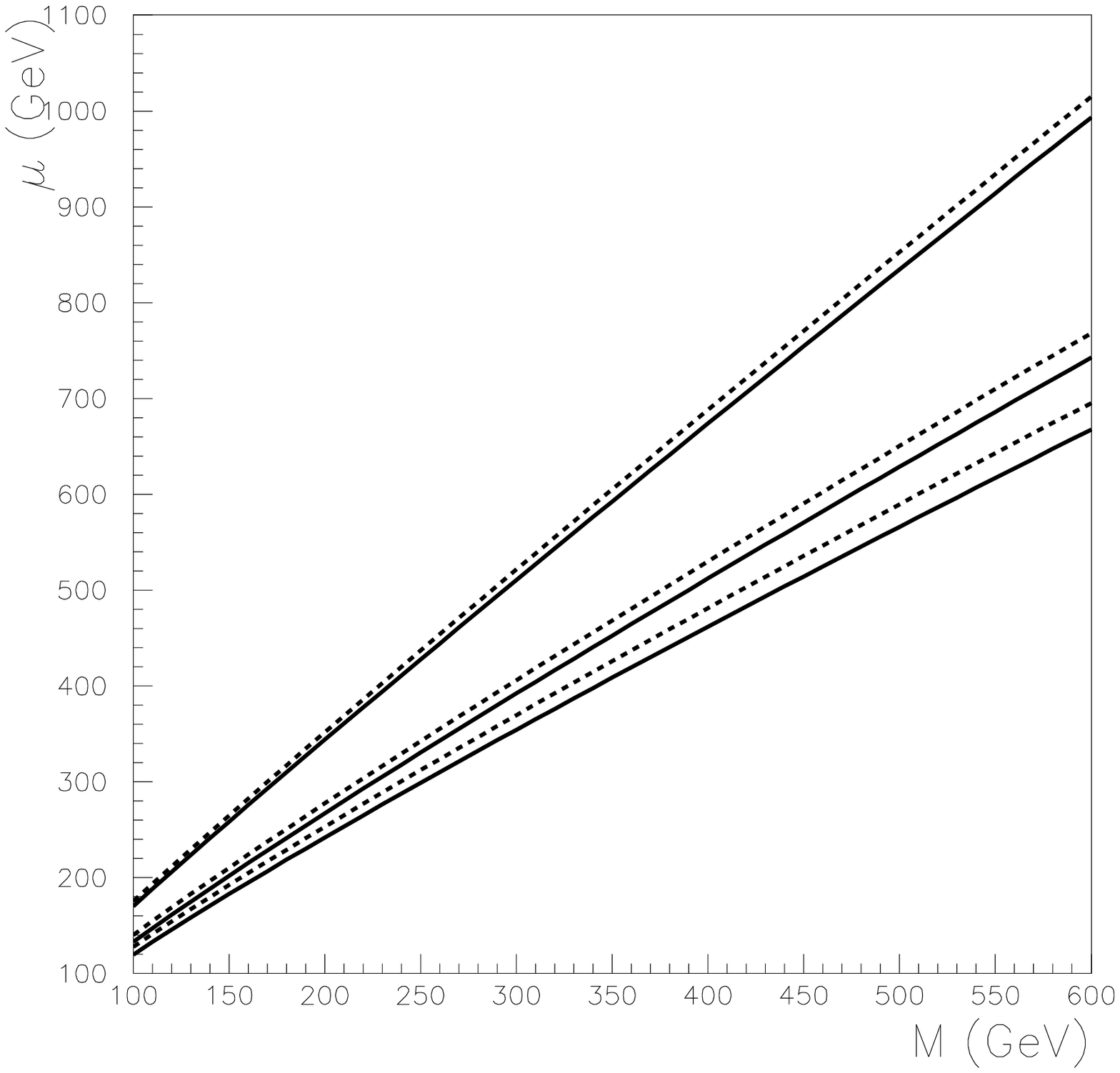}
\vskip 0.3cm
\hskip 3.8cm
a)
\hskip 7cm
b)
\caption{\footnotesize The dependence of a) $m_h$ and b) $\mu$ on $M$
for three different choices of $A=+2 M, 0, -2 M$ (from top to bottom).
Here we fix $m$=100 GeV for definiteness.  The dotted and solid lines
are for $\tanb = 5, 10$ respectively.}
\end{figure}

Changing the sign of $\mu$ from positive to negative has both indirect
and direct effects. Since now $A$ and $\mu$ appear with opposite sign
in the $\tilde{t}_L - \tilde{t}_R$ mixing terms, the predicted value
of $m_h$ is reduced slightly. One thus has to increase the minimal
allowed value of $M$ in order to satisfy the bound $m_h > 111$ GeV. In
fact, we didn't find any allowed solutions within the scanning range
for $\tanb =4, \ A_0 = 0$ and $\mu < 0$, while the minimal allowed
value of $M$ for $\tanb=10$ is increased by $\sim 40$ GeV, leading to
a reduction of the maximal possible $|\Delta a_\mu|$ to about $2.4
\times 10^{-9}$. Note that for fixed values of the other parameters,
the magnitude of $\Delta a_\mu$ is almost independent of the sign of
$\mu$, but the sign of $\Delta a_\mu$ changes when the sign of $\mu$
is flipped. In contrast, the sizes of the $\lsp \lsp (h,H)$ couplings
do depend quite significantly on the sign of $\mu$, unless $\tanb \gg
1$ \cite{dn3}. In particular, for $\mu < 0$ strong cancellations occur
\cite{dn5} both within different contributions to the same coupling,
and between the $h$ and $H$ exchange contributions to $\sig$. As a
result, for $\tanb = 10$, cross sections as low as $4 \times 10^{-11}$
pb are possible in the cosmologically favored region, and the maximal
allowed value of $\sig$ is reduced by more than one order of
magnitude, to $5 \times 10^{-10}$ pb. The importance of the sign of
$\mu$ diminishes with increasing \tanb; hence results for $\tanb=30$
and $\mu < 0$ are quite similar to the results for $\tanb=30$ shown in
Fig.~1. Finally, we mention that in models with universal scalar
masses, solutions with $\mu < 0$ tend to have too large a branching
ratio for radiative $b \rightarrow s \gamma$ decays
\cite{bsg}.\footnote {Recently the leading higher order corrections to
this branching ratio have been computed \cite{bsgnlo}. While a
detailed parameter scan has not yet been performed, the published
results indicate that scenarios with degenerate scalar masses and $\mu
< 0$ remain problematic at large \tanb. On the other hand, for $\mu >
0$ agreement with the data can be obtained even for $\tanb \simeq
45$.} However, this prediction is sensitive to details of the flavor
structure of the soft breaking terms, unlike the quantities we have
depicted in Fig.~1. We therefore do not attempt to analyze the
constraint from $b \rightarrow s \gamma$ decays quantitatively.

\section{More general models}

In this section we relax our assumptions, allowing for non--universal
soft scalar masses at the GUT scale, while keeping the unification of
the gaugino masses. As specific models, we consider a string--inspired
supergravity model and an $SO(10)$ Grand Unified model.

Soft SUSY breaking terms have been studied within the framework of
string--inspired supergravity models \cite{string-SG}. In a wide class
of models one obtains the following relations among soft SUSY breaking
terms:
\begin{eqnarray}
 A_{ijk} = -M, \qquad 
 m_i^2 + m_j^2 + m_k^2 = M^2,
\label{string-sum}
\end{eqnarray}
for non--vanishing Yukawa couplings $Y_{ijk}$ of chiral superfields
$\Phi^i$, $\Phi^j$ and $\Phi^k$. These relations hold if the theory is
target--space duality--invariant and the Yukawa couplings in the
supergravity basis are field--independent. In addition we assume that
only the $F-$terms of the dilaton and the moduli superfields
contribute to SUSY breaking, and that the vacuum energy vanishes. The
generic form of supergravity theories leading to these relations has
been obtained in Ref.~\cite{sum-rule}.\footnote{These relations are
RG--invariant at one-loop for a single gauge group.}  These relations
allow non--universality of soft scalar masses, but their overall
magnitudes are bounded by the gaugino mass through the sum. This point
is significant.

Here we apply the relations (\ref{string-sum}) to the up and down
sectors of (s)quarks and the (s)lepton sector separately. To be
explicit, we have the relations between the soft breaking
contributions to the squared scalar masses,
\begin{eqnarray} \label{string}
m_Q^2 + m_U^2 + m_{H_u}^2 = M^2 \nonumber\\
m_Q^2 + m_D^2 + m_{H_d}^2 = M^2  \nonumber\\
m_L^2 + m_E^2 + m_{H_d}^2 = M^2,
\end{eqnarray}
and a universal $A$-term, $A=-M$.

In fact, these relations include quite a large parameter space, where
sfermion masses are not equal to each other.  However, as the first
trial we take universal sfermion masses; this remains the simplest
solution of the supersymmetric flavor problem. We shall later comment
on a case with non--universal sfermion masses.

The $SO(10)$ theory incorporates a complete generation of MSSM matter
superfields into the 16--dimensional spinor representation,
$\Psi_{16}$. In addition to these matter superfield, the minimal
$SO(10)$ model includes a 10--dimensional Higgs superfield $\Phi_{10}$
which contains the two Higgs superfields of the MSSM (as well as their
$SU(3)$ triplet, $SU(2)$ singlet partners). When $SO(10)$ breaks to
the MSSM gauge group $SU(3)_C \times SU(2)_L \times U(1)_Y$,
additional $D-$term contributions (parameterized by $M_D^2$ which can
be either positive or negative) to the soft SUSY breaking masses arise
\cite{dterm}:
\begin{eqnarray} \label{so10}
m^2_Q = m^2_E = m^2_U = m^2_{16} + M^2_D \nonumber \\
m^2_D = m^2_L = m^2_{16} - 3 M^2_D, ~ 
m^2_{H_{u,d}} = m_{10}^2 \mp 2 M^2_D,
\end{eqnarray}
%
where $m_{16}$ and $m_{10}$ are scalar soft breaking masses for fields
in the {\bf 16} and {\bf 10} dimensional representations of
$SO(10)$, respectively.

The modifications (\ref{string}) and (\ref{so10}) of the mSUGRA
boundary conditions change our predictions in two different ways:
through modifications of the slepton spectrum (which change $\Delta
a_\mu$ and $\Omega_\chi h^2$), and through the changed value of
$|\mu|$ at the weak scale, which mostly affects $\sig$. While the
changes of the slepton spectrum can be read off directly from
eqs.(\ref{string}) and (\ref{so10}), the changes of $|\mu|$ are more
subtle. The main effects can be understood as follows. In the mSUGRA
scenario, the contributions of $m$ and $M$ to the weak scale values of
the soft breaking Higgs boson masses can be parameterized as
\begin{eqnarray} \label{msugra}
m^2_{H_d} \simeq m^2 + 0.5 M^2; \nonumber \\
m^2_{H_u} \simeq \epsilon_H m^2 - 3.3 M^2 ,
\end{eqnarray}
where we have assumed $\sin\beta \simeq 1$ but ignored contributions
from the bottom Yukawa coupling. The coefficient $\epsilon_H$ is
small, because the GUT scale value of $m^2_{H_u}$ is canceled by the
scalar masses appearing in the RG running.\footnote{The effective
$\epsilon_H$ is slightly negative for low SUSY breaking scale, but
turns positive if this scale is large. Recall that the relevant scale
for the analysis of gauge symmetry breaking increases with increasing
sparticle masses.} The effect of non--universality on $m_{H_u}^2$ can
be parameterized by introducing $m_Q^2 + m_U^2 + m_{H_u}^2 =3 m_s^2$
and $\delta m_H^2 = m_{H_u}^2 - m_s^2$:
\begin{equation} \label{nonuniv}
m_{H_u}^2 = \delta m_H^2 + \epsilon_H m_s^2 - 3.3 M^2.
\end{equation}
This equation follows because the radiative correction to $m^2_{H_u}$
is proportional to $m^2_Q + m^2_U + m^2_{H_u}$, hence the effect of
RGE running is the same as in mSUGRA with the replacement $m^2
\rightarrow m_s^2$.  If $\sin \beta \simeq 1$, correct gauge symmetry
breaking requires $\mu^2 \simeq -m^2_{Hu} -M_Z^2/2$, where all
quantities are taken at the weak scale. Hence $\delta m_H^2$ directly
affects the value of $|\mu|$ at the weak scale. For the string
scenario (\ref{string}) the effect is limited because the total scalar
mass is bounded from above by the gaugino mass, but for the $SO(10)$
scenario (\ref{so10}) the change of $|\mu|$ from its mSUGRA prediction
can be more drastic.

Fig.~3 shows the correlation between $\Delta a_\mu$ and $\sig$ for
$\tanb = 10$ and $\mu > 0$ in three different models: mSUGRA, the
string scenario described by eq.(\ref{string}) with universal sfermion
masses ($\equiv m$), and the $SO(10)$ GUT model. Note that the
assumption of universal sfermion masses implies $m_{H_d}^2 =
m_{H_u}^2$ at the GUT scale in the stringy model; the value of the
Higgs soft breaking masses is determined by eq.(\ref{string}) for
given $m$ and $M$, and $A=-M$. For the $SO(10)$ GUT model, we assume
$m_{16} = m_{10} \equiv m$ and fix $A=0$ and $M_D^2 = -50000$
GeV$^2$. As before, we scan the region $m < 500$ GeV and $M < 600$
GeV, requiring the physical Higgs mass and chargino mass to lie above
their experimental lower bounds, and excluding regions with charged
LSP. The heavily marked points again indicate the cosmologically
favored region where $\Omega_\chi h^2 < 0.3$.
\begin{figure}[htbp]
\begin{center}
\includegraphics[width=8.5cm,angle=0]{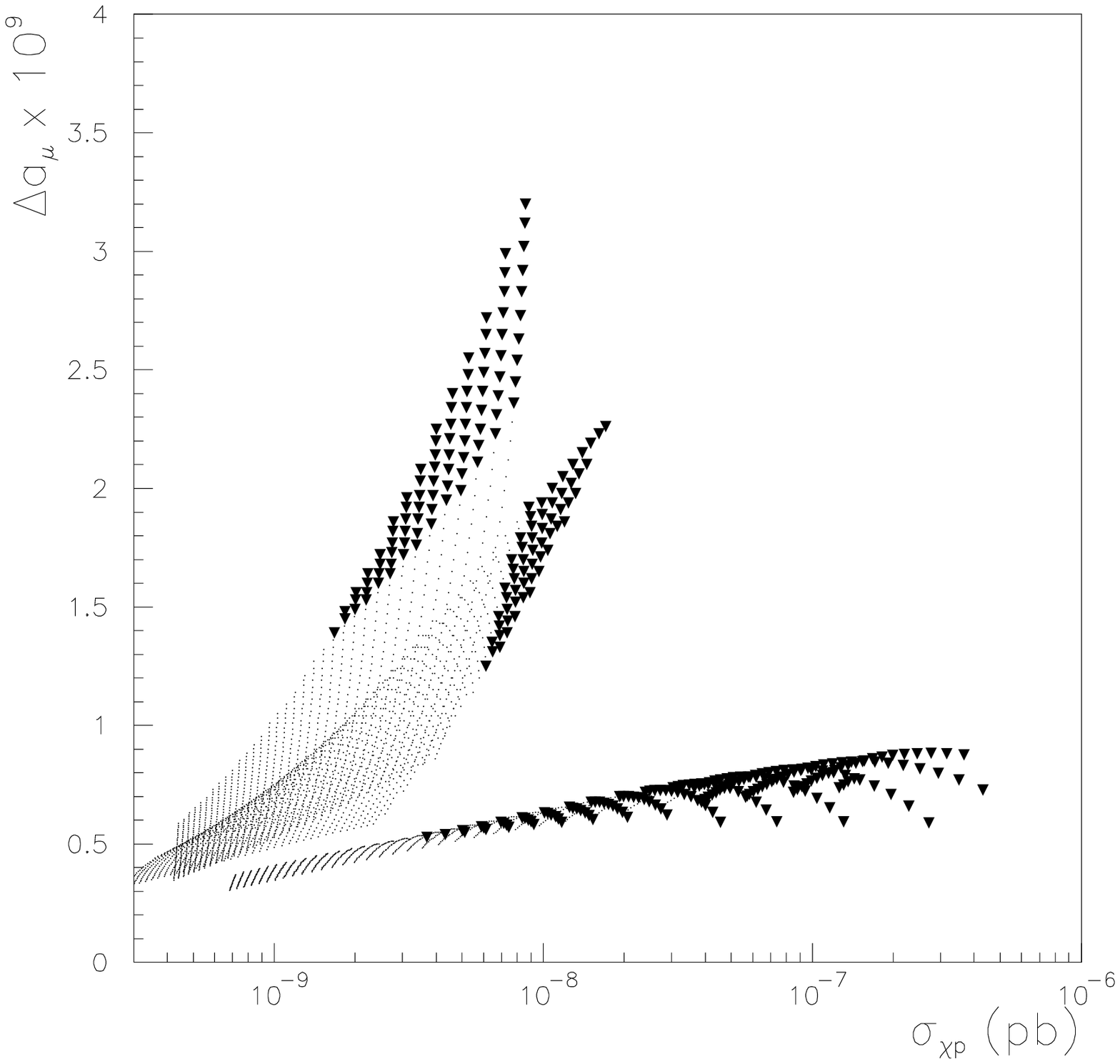}
\end{center}
\caption{\footnotesize $\Delta a_\mu$ vs. $\sigma_{\chi p}$ in three
different models : mSUGRA, a superstring scenario and the minimal
$SO(10)$ GUT model. We chose $\tanb = 10$ and $\mu > 0$. The scanning
procedure and notation is as in Fig.~1.}
\label{fig2}
\end{figure}

In our superstring scenario the higgsino component of the LSP can be
larger or smaller than in mSUGRA, depending on the ratio $m/M$.  For
fixed $M$, a small $m$ gives a larger higgsino component of the LSP
because then $m_{H_{u,d}}^2$ should be larger than $m^2$ in order to
satisfy eq.(\ref{string}), i.e. $\delta m_H^2 > 0$ in
eq.(\ref{nonuniv}). This in turn reduces the supersymmetric
contribution $\mu^2$ to the Higgs mass term which is required to
obtain the correct $Z$ boson mass. On the other hand, for $m^2 >
M^2/3$, eq.(\ref{string}) gives $\delta m_H^2 < 0$, leading to even
larger $|\mu|$ than in mSUGRA. Altogether this model allows for a
broader range of values of $\sig$, with maximum around $1.7 \times
10^{-8}$ pb, about a factor of 2 above the maximal mSUGRA
prediction. On the other hand, the maximum value of $\Delta a_\mu$ is
reduced to $2.3 \times 10^{-9}$ from $3.3 \times 10^{-9}$ in
mSUGRA. This reduction mainly comes from the fact that the minimal
allowed value of $M$ is raised by about 40 GeV due to the choice
$A=-M$ (rather than $A=0$ in the mSUGRA case); we saw in Fig.~2 that
negative values of $A$ tend to reduce $m_h$ if all other input
parameters are kept fixed. The variation of $\mu$ is less important
for $\Delta a_\mu$. The reason is that increasing $|\mu|$ increases
the higgsino mass and reduces higgsino--gaugino mixing, which reduces
$|\Delta a_\mu|$, but at the same time increases $\tilde{\mu}_L -
\tilde{\mu}_R$ mixing, which tends to increase $|\Delta a_\mu|$.

We also performed a scan (not shown) for a second ansatz for scalar
soft breaking masses which is compatible with the general expressions
(\ref{string}). Here we kept the soft breaking contributions to the
squared squark and Higgs boson masses equal to each other. This
implies $m^2_Q = m^2_U = m^2_D = m^2_{H_u} = m^2_{H_d} = M^2/3$ and
$m^2_E + m^2_L = 2 M^2/3$. We then allowed $m_E$ and $m_L$ to differ
from each other, keeping universality among different generations.
This scan gives results for $\sig$ which are quite similar to those of
the mSUGRA scenario; however, the predictions for $a_\mu$ are
generally lower than in mSUGRA. This is true in particular for
parameter values that satisfy $\Omega_\chi h^2 < 0.3$. For the given
relation between average slepton mass and gaugino mass, an acceptable
value of the relic density can only be achieved\footnote{We note in
passing that the dilaton--dominated scenario \cite{dilaton}, where all
scalar soft breaking masses are equal to $M/\sqrt{3}$, cannot
simultaneously accommodate $m_h > 111$ GeV and $\Omega_\chi h^2 <
0.3$, unless \tanb\ is very large.} if $m_E < m_L$; note that due to
its larger hypercharge, $\tilde{e}_R$ exchange contributes 8 times
more strongly to the annihilation of bino--like LSPs than
$\tilde{e}_L$ plus $\tilde{\nu}$ exchange does. This mass splitting
between $\tilde{e}_R$ and $\tilde{\nu}$ reduces $\Delta a_\mu$, since
the (usually dominant) sneutrino--chargino loop contributions are
suppressed compared to the scenario with universal slepton masses. The
main difference to our first stringy scenario, where we allowed
different soft breaking contributions for sfermions and Higgs bosons
while keeping sfermion masses equal to each other, are the lower
maximally allowed values of both $\sig$ and $\Delta a_\mu$. This also
leads to smaller values of $\Delta a_\mu$ and $\sig$ in the
cosmologically acceptable region: the minimal $\sig$ which is
compatible with $\Omega_\chi h^2 < 0.3$ is reduced to $\sim 2.5 \times
10^{-9}$ pb from $\sim 6 \times 10^{-9}$ pb, while the smallest
cosmologically allowed $\Delta a_\mu$ is reduced from $\sim 1.25
\times 10^{-9}$ to $\sim 0.9 \times 10^{-9}$.

While the predictions of these versions of the superstring inspired
model differ from those of mSUGRA ``only'' by about a factor of 2, the
$SO(10)$ GUT model with $M_D^2 = -50000$ GeV$^2$ leads to a
dramatically changed correlation between $\Delta a_\mu$ and
$\sig$. Since $m^2_{H_u}$ at the GUT scale is larger than in mSUGRA,
the higgsino component of the LSP increases. This effect is especially
significant for small $M$, where the gaugino mass contributions to the
RGE of the scalar masses are not so large, so that a modification of
the GUT scale boundary condition more strongly affects weak scale
values of scalar masses; see eq.(\ref{nonuniv}). In this region $\sig$
increases quite a lot, reaching values up to $\sim 4 \times 10^{-7}$
pb, which could be detected in the near future. In contrast, the
allowed values of $\Delta a_\mu$ are less than $1 \times 10^{-9}$,
well below near future sensitivity. This reduction is caused by the
increase of the sneutrino mass and the decrease of the $\tilde{\mu}_R$
mass due to the $D-$term contributions in eq.({\ref{so10}). Because of
this large slepton mass splitting, eq.(\ref{amu}) is not applicable in
this scenario.  The $D-$terms suppress the contribution from
chargino--sneutrino loops, while enhancing the contribution from
neutralino--smuon loops. These two contributions can now have
comparable magnitude, but opposite sign, leading to significant
cancellations in the small $m$ region. These cancellations explain why
we sometimes find an anti--correlation between $\sig$ and $\Delta
a_\mu$ in this model, i.e. decreasing $m^2$ can lead to decreasing
$\Delta a_\mu$ (and increasing $\sig$, as usual). Of course, for large
$m$ all contributions to $\Delta a_\mu$ are small.
\begin{figure}[htbp]
\begin{center}
\includegraphics[width=8.5cm,angle=0]{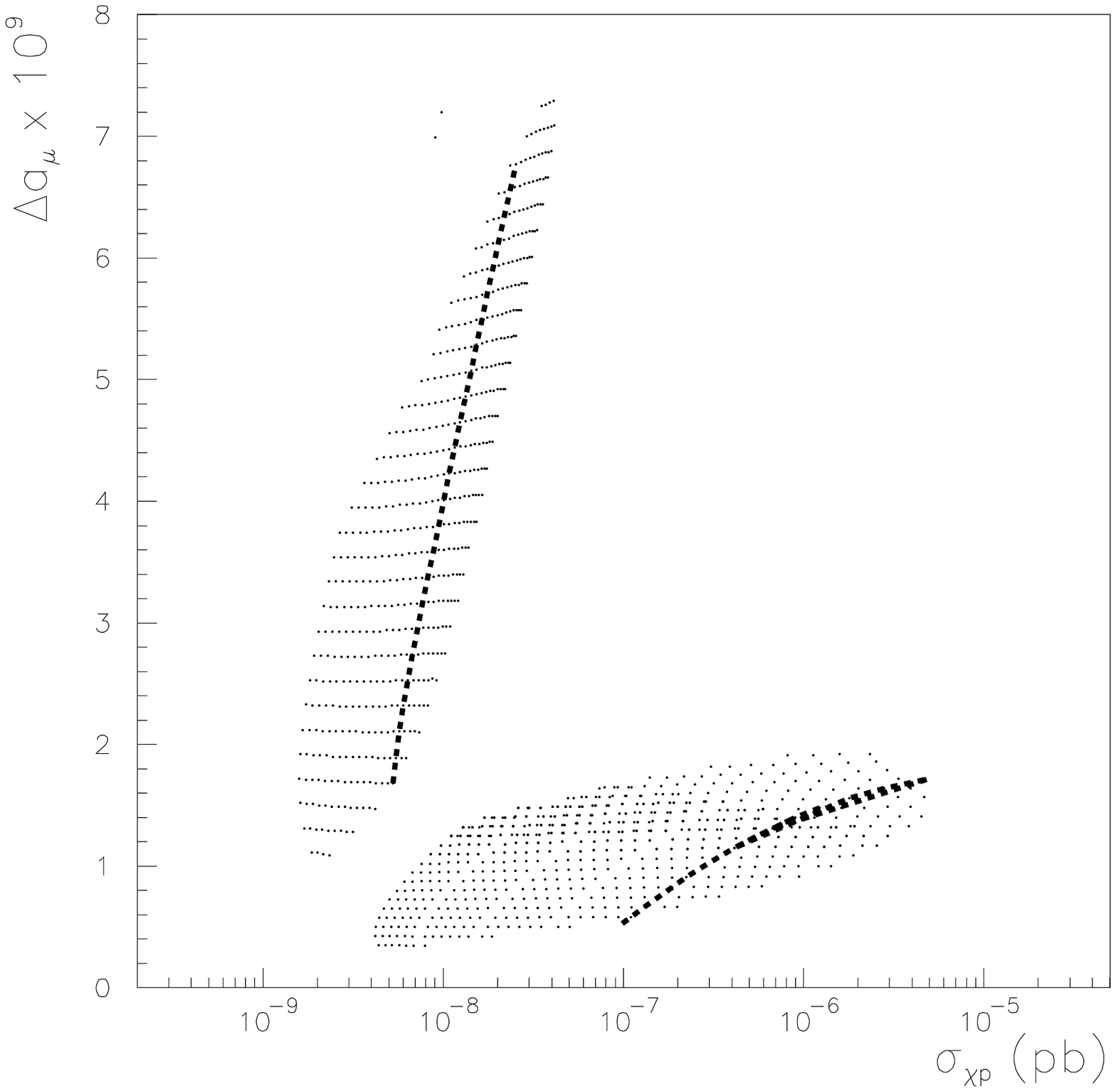}
\end{center}
\caption{\footnotesize $\Delta a_\mu$ vs. $\sig$ 
in mSUGRA and the $SO(10)$ model with $M_D^2=-50000$ GeV$^2$. 
Here, we fix $m$ and $M$ such that $\mlsp \simeq 120$ GeV and 
$\mer \simeq 182$ GeV, and vary $A$ and $\tan\beta$.}
\label{fig4}
\end{figure}

In Fig.~3 the models with and without $SO(10)$ $D-$terms are well
separated. Therefore one might conclude that one can test for the
existence of $SO(10) \ D-$terms by measuring $\Delta a_\mu$ and/or
$\sig$. However, in Fig.~3 we fixed the values of $\tanb, A$ and
$M_D^2$. In order to see the effects of varying \tanb\ and $A$, we fix
$m= 140$ GeV and $M = 280$ GeV in mSUGRA, and $m=260$ GeV and $M =
300$ GeV in the $SO(10)$ model, so that $\mlsp \simeq 120$ GeV and
$\mer \simeq 182$ GeV in both cases. We then scan the region of $-2 M
< A < 2 M$ and the entire allowed range of \tanb. Both the lower and
upper bounds on \tanb\ come from Higgs searches at LEP; recall that in
models with radiative symmetry breaking, {\em all} neutral Higgs
bosons become quite light as \tanb\ approaches its upper bound
\cite{dn1}. Fig.~4. shows the result of this scan. Here the short--dashed
line corresponds to $A = 0$, with positive (negative) values of $A$
falling to the left (right) side of this line.  As expected, $\Delta
a_\mu$ increases when \tanb\ increases, and depends only weakly on
$A$.  On the other hand, $\sig$ increases with increasing \tanb, but
decrease when $A$ is increased. As noted earlier, this reduction of
$\sig$ comes from the fact that the Higgs masses increase, and the LSP
becomes more bino--like, when $A$ increases.

In this figure, mSUGRA scenarios are characterized by relatively large
$\Delta a_\mu$ and small $\sig$, but the opposite is true for the
$SO(10)$ GUT model with $M_D^2$ = -50000 GeV$^2$ in most of the
allowed region. The two classes of models thus remain quite well
separated. Of course, it is by no means certain that the extra
$D-$term contribution in the $SO(10)$ model should be large and
negative. In fact, if one requires that all three third generation
Yukawa couplings unify at the GUT scale, which is not unreasonable in
minimal $SO(10)$ since all third generation superfields reside in a
single representation of $SO(10)$, consistent radiative symmetry
breaking is only possible if $M_D^2 > 0$ \cite{baertata}. In
Figs.~5a,~b we therefore show the dependence of $\Delta a_\mu$ and
$\sig$ on $M_D^2$, for $m=260$ GeV, $M=300$ GeV, $A=0$ and
$\tanb=10$. The upper and lower bounds on $M_D^2$ come from the
requirement that $SU(2)$ singlet and doublet sleptons, respectively,
should be heavier than the lightest neutralino.  We see that $\sig$
increases very quickly for negative $M_D^2$; near the lower bound on
$M_D^2$ one even approaches the sensitivity of {\em current} Dark
Matter search experiments. Conversely, $\Delta a_\mu$ increases if
$M_D^2$ is raised above 0, but the maximal increase amounts to less
than a factor of 2 compared to the mSUGRA situation, which corresponds
to $M_D^2=0$. Note, however, that Yukawa unification also requires
$\tanb \gsim 35$ \cite{baertata}; the combined enhancement of $\Delta
a_\mu$ from large \tanb\ and from $M_D^2 > 0$ implies that
measurements of the anomalous dipole moment of the muon might be one
of the best ways to test $SO(10)$ models with Yukawa unification.

\begin{figure}[htbp]
\hskip 0.0cm
\includegraphics[width=8.5cm,angle=0]{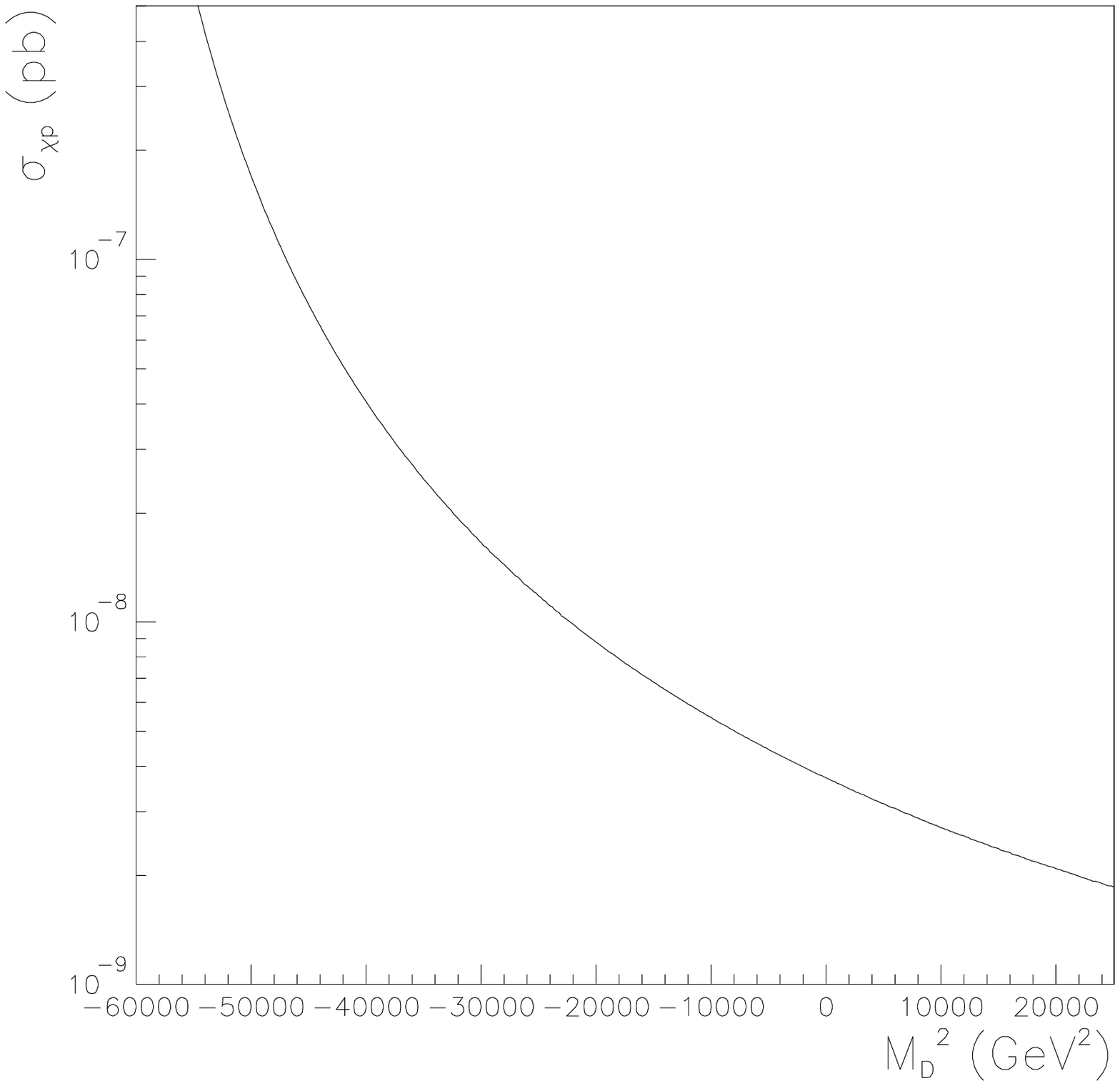}
\hskip 0.5cm
\includegraphics[width=8.5cm,angle=0]{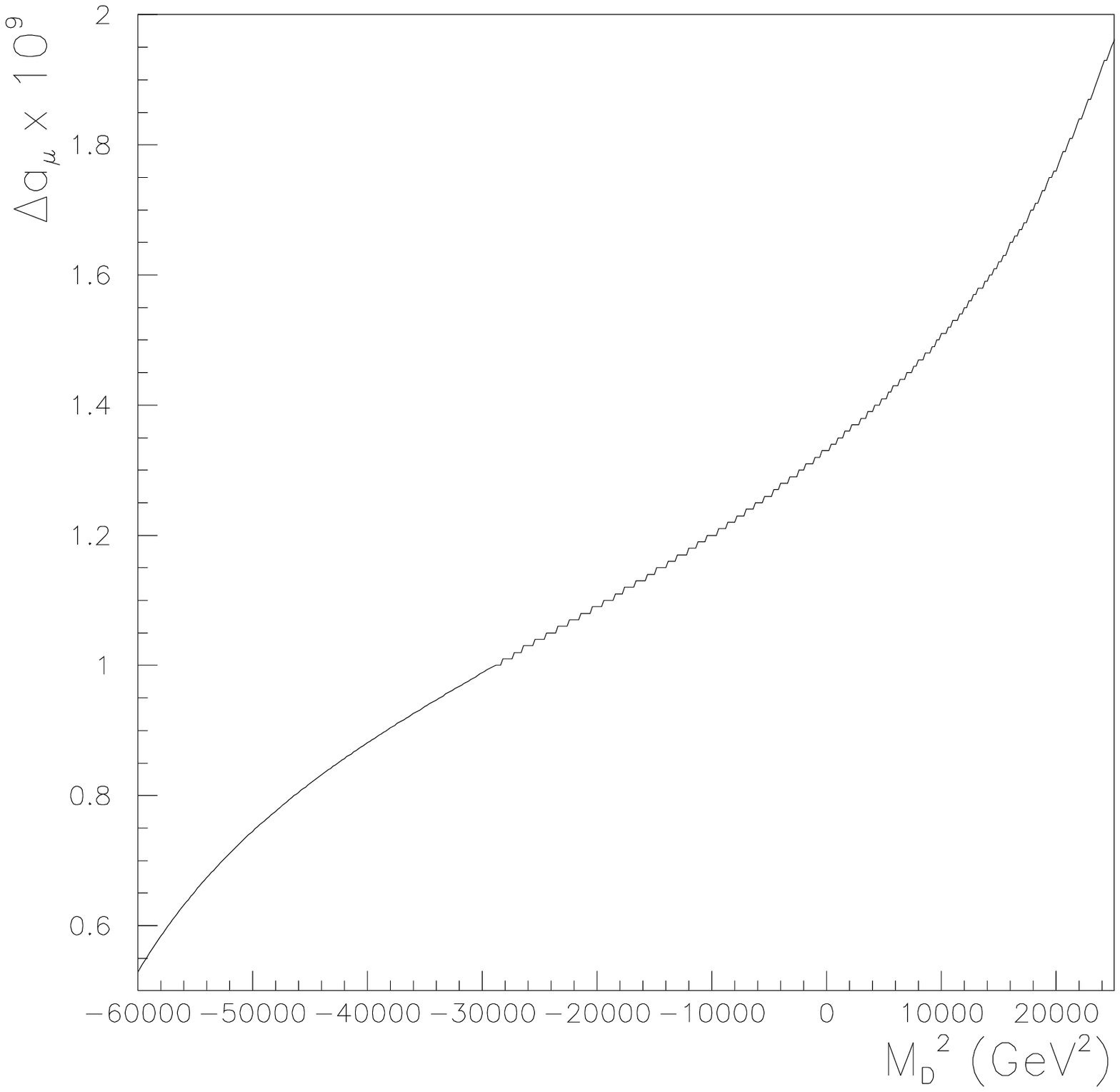}
\vskip 0.3cm
\hskip 3.8cm
a)
\hskip 7cm
b)
\caption{\footnotesize The dependence of a) $\sig$ and b) $\Delta
a_\mu$ on $M^2_D$ in the $SO(10)$ model, for $m=260$ GeV, $M=300$ GeV,
$A=0$ and $\tanb=10$.}
\end{figure}
\label{fig5}

\section{Conclusions}

In this paper, we investigated the SUSY contribution to the muon MDM
$a_\mu$, and to the spin--independent neutralino--proton cross section
$\sig$, in several SUSY models with either universal or non--universal
soft scalar masses at the GUT scale. Both these observables can become
significant if \tanb\ is large, but will be difficult to measure in
the small \tanb\ region. Both quantities are sensitive to chirality
violation in the matter (s)fermion sector, which in the MSSM is
enhanced for large \tanb. Conversely, if \tanb\ is small the
experimental Higgs mass bounds tightly restrict the allowed SUSY
parameter space, forcing the SUSY breaking scale to be quite
high. While these general statements are fairly model--independent as
long as one sticks to the field content of the MSSM, quantitative
predictions do depend significantly on details of the spectrum of
superparticles, in particular on the implementation of radiative gauge
symmetry breaking.

We considered three different models: mSUGRA, a superstring scenario
where SUSY is broken by the $F-$terms of the dilaton and the moduli
superfields, and an $SO(10)$ GUT model. The superstring model in
principle has a very large parameter space, since it allows for
non--universal masses for matter sfermions and Higgs bosons, subject
only to the sum rule (\ref{string-sum}). Here we explored two
orthogonal directions in this parameter space, where universality is
violated either only in the Higgs sector or only in the slepton
sector. We found that allowing the Higgs masses to differ from the
sfermion masses at the GUT scale can vary the predicted value of
$\sig$ by a factor of a few in either direction, compared to the
mSUGRA prediction for the same parameters. Given the uncertainty in
the prediction of $\sig$ due to unknown hadronic matrix elements, and
the uncertainty in the predicted LSP--nucleus scattering rate due to
the uncertainty of the ambient LSP flux, such a variation is barely
significant. The prediction of $\Delta a_\mu$ is almost insensitive to
this variation of the Higgs soft breaking masses. Conversely, if we
maintain the universality of the soft breaking contributions to squark
and Higgs masses but allow the masses of $SU(2)$ doublet and singlet
sleptons to differ, we can change the prediction for $\Delta a_\mu$ by
a factor of $\sim 2$ around the mSUGRA value, while keeping $\sig$
constant; the predicted LSP relic density, and hence the
cosmologically preferred region of parameter space, also depends on
this ratio of slepton masses.  Generally speaking, in mSUGRA as well
as its stringy variant the measurement of the anomalous dipole moment
of the muon seems to hold more promise in the near future than
searches for LSP Dark Matter; this is largely due to the most recent
severe constraints on the Higgs sector from LEP experiments.

We should caution the reader that this conclusion crucially depends on
our choice to only study scenarios where both squarks and gluinos can
be produced at the LHC. Given current slepton and chargino search
limits, this implies that sfermion and gaugino masses are very roughly
of the same order of magnitude. If one allows scalar masses $m \gsim
1$ TeV and $m^2 \gg M^2$, one can find solutions in mSUGRA where the
LSP has a large, or even dominant, higgsino component
\cite{focus}. Since in this region of parameter space all sfermions
are very heavy, the supersymmetric contribution to $a_\mu$ is totally
negligible, but $\sig$ can be very large \cite{focus2}.

Much more dramatic deviations from mSUGRA predictions are possible if
one introduces $SO(10)$ $D-$term contributions to scalar masses. In
this model the maximal allowed value of $\sig$ for given \tanb\ can
exceed the mSUGRA prediction by more than a factor of 30, if the
$D-$term contribution to the mass of the Higgs boson that couples to
the top quark is positive. For the same sign of the $D-$terms the
supersymmetric contribution to $a_\mu$ is reduced by a factor of a
few, since the change of the slepton masses increases the importance
of neutralino loop diagrams compared to chargino loops, leading to a
strong cancellation in the total result. In this kind of model direct
Dark Matter searches therefore appear to be more promising in the next
few years than measurements of $a_\mu$.

A striking result of our analysis is that even if we restrict squark
and gluino masses to lie within discovery range of the LHC, and in
addition fix the value of \tanb, the prediction for $\sig$ still
varies by almost three orders of magnitude within the range of models
we studied; in contrast, the prediction for $\Delta a_\mu$ ``only''
varies by about one order of magnitude; see Fig.~3. One reason for
this difference is that $\sig$ measures squared Feynman amplitudes,
while $a_\mu$ is directly proportional to a (loop) amplitude. In
addition, we saw that $\Delta a_\mu$ does not depend on $\mu$ very
sensitively, since various effects tend to cancel. In contrast, the
sensitivity of $\sig$ to $\mu$ is often enhanced by cancellations
between diagrams with different $\mu-$dependence. A similar remark
holds for the \tanb\ dependence. Fig.~1 shows that it is ``only''
linear for $\Delta a_\mu$, but much stronger for $\sig$. In this case
the strong sensitivity of $\sig$ is partly due to the dependence of
the mass of the heavy CP--even Higgs boson on \tanb, which is a direct
consequence of radiative gauge symmetry breaking. Given the
theoretical uncertainties involved in predicting Dark Matter detection
rates, this strong parameter dependence of $\sig$ is in some sense
fortunate; it means that even an order--of--magnitude determination of
$\sig$ will allow us to make significant statements about
supersymmetric parameters. Given its comparatively mild parameter
dependence, it is also fortunate that the prediction of $a_\mu$ is
much cleaner, if we assume that the hadronic uncertainty can be
reduced to the level of the expected experimental uncertainty, which
seems feasible.

Our overall conclusion is that measurements of $a_\mu$ and/or $\sig$
might well yield a positive signal in the next few years. These
measurements can help to distinguish between different supersymmetric
models. Since well motivated models exist where one of these
quantities is sizable while the other is small, it is important to
measure, or at least constrain, {\em both} of them as precisely as
possible. On the other hand, we also saw that two models with very
different scalar spectrum (the $SO(10)$ model and mSUGRA with $m^2 \gg
M^2$) can lead to very similar predictions for both $a_\mu$ and
$\sig$. This once again illustrates the truism that indirect
measurements can provide crucial information, but they can never
completely replace direct searches for new particles.

\subsection*{Acknowledgements}
The work of M.D. was supported in part by the ``Sonderforschungsbereich
375--95 f\"ur Astro--Teilchenphysik'' der Deutschen
Forschungsgemeinschaft. M.M.N. was supported in part by a Grant-in-Aid 
for Scientific Research from the Ministry of Education (12047217).


\begin{thebibliography}{50}          
\bibitem{SUSY}
For a review, see H.~E.~Haber and G.~L.~Kane, Phys. Rep. {\bf 117}, 75
(1985).

\bibitem{witten}
E. Witten, Nucl. Phys. B{\bf 188}, 513 (1981).

\bibitem{amaldi}
J. Ellis, S. Kelley and D.V. Nanopoulos, Phys. Lett. B{\bf 249}, 441
(1990), and Phys. Lett. B{\bf 260}, 131 (1991);
C. Giunti, C.W. Kim and U.W. Lee, Mod. Phys. Lett. A{\bf 6}, 1745
(1991); 
U. Amaldi, W. de Boer and H. F\"urstenau, Phys. Lett. B{\bf 260}, 447
(1991); P. Langacker and M. Luo, Phys. Rev. D{\bf 44}, 817 (1991).

\bibitem{colrev}
See e.g. ATLAS Collab., 
ATLAS Detector and Physics Performance Technical Design Report. 
CERN/LHCC99-19.\\
http://atlasinfo.cern.ch/Atlas/GROUPS/PHYSICS/TDR/access.html;
Proceedings of the ``1996 DPF/DPB Summer Study on High-Energy
Physics''.  Ed. D.~G.~Cassel, L.~T.~Gennari and R.~H.~Siemann;
E.~Accomando {\it et al.}, [ECFA/DESY LC Physics Working Group
Collaboration], Phys.\ Rep.\  {\bf 299}, (1998),
hep--ph/9705442.

\bibitem{jungman}
For a review, see G. Jungman, M. Kamionkowski and K. Griest,
Phys. Rep. {\bf 267}, 195 (1996).

\bibitem{DAMA} 
DAMA Collaboration, Phys. Lett. B{\bf 480}, 23 (2000).

\bibitem{CDMS1} 
CDMS Collaboration, Nucl. Instrum. Meth. A{\bf 444}, 345 (2000),
astro--ph/0002471. 

\bibitem{CDMS2}
R. Schnee for the CDMS collab., talk at {\it DARK 2000}, Heidelberg,
Germany, July 2000.

\bibitem{cresst} 
CRESST Collaboration, Astropart. Phys. {\bf 12}, 107 (1999),
hep--ex/9904005.

\bibitem{carey} 
R. Carey, plenary talk given at ICHEP2000; see
http://ichep2000.hep.sci.osaka-u.ac.jp/scan/0801/pl/carey/index.html

\bibitem{e821} 
E821 Collaboration, Nucl. Phys. Proc. Suppl. {\bf 76}, 253 (1999)

\bibitem{kino}
V.W. Hughes and T. Kinoshita, Rev. Mod. Phys. {\bf 71}, D133 (1999). 

\bibitem{amugen}
The supersymmetric contribution to $a_\mu$ has previously been studied
by numerous authors. A selection includes:
J. Ellis, J. Hagelin and D.V. Nanopoulos, Phys. Lett. {\bf 116}B, 283 (1982);
J.A. Grifols and A. Mendez, Phys. Rev. {\bf D26}, 1809 (1982); 
R. Barbieri and L. Maiani, Phys. Lett. {\bf B117}, 203 (1982); 
D.A. Kosower, L.M. Krauss and N. Sakai, Phys. Lett. {\bf B133}, 305
(1983); 
T.C. Yuan, R. Arnowitt, A.H. Chamseddine and Pran Nath, Z. Phys. {\bf
C26}, 407 (1984);
H. K\"onig, Mod. Phys. Lett. {\bf A7}, 279 (1992);
U. Chattopadhyay and Pran Nath, Phys. Rev. {\bf D53}, 1648 (1996),
hep--ph/9507386;
M. Carena, G.F. Giudice and C.E.M. Wagner, Phys. Lett. {\bf B390}, 234
(1997), hep--ph/9610233.

\bibitem{dn3}
M.~Drees and M.~M.~Nojiri,  Phys. Rev. {\bf D47}, 376 (1993).

\bibitem{dn5}
M. Drees and M.M. Nojiri, Phys. Rev. {\bf D48}, 3483 (1993),
hep--ph/9307208.

\bibitem{moroi} T. Moroi, Phys. Rev. D{\bf 53}, 6565 (1996); Erratum-ibid.
{\bf D56}, 4424 (1997), hep--ph/9512396.

\bibitem{sigold}
Early analyses of neutralino--nucleon scattering include:
M.W. Goodman and E. Witten, Phys. Rev. {\bf D31}, 3059 (1985);
K. Griest, Phys. Rev. {\bf D38}, 2357 (1988), Erratum-ibid. {\bf D39},
3802, (1989);
M. Srednicki and R. Watkins, Phys. Lett. {\bf B225}, 140 (1989);
M. Kamionkowski, Phys. Rev. {\bf D44} 3021 (1991);
G.B. Gelmini, P. Gondolo and E. Roulet, Nucl. Phys. {\bf B351}, 623
(1991);
J. Ellis and R.A. Flores, Phys. Lett. {\bf B263}, 259 (1991), and
Phys. Lett. {\bf B300}, 175 (1993).

\bibitem{signew}
Recent analyses of neutralino--nucleon scattering in the MSSM include:
H. Baer and M. Brhlik, Phys. Rev. {\bf D57}, 567 (1998),
hep--ph/9706509; 
E. Accomando, R. Arnowitt, B. Dutta and Y. Santoso, Nucl. Phys. {\bf
B585}, 124 (2000), hep--ph/0001019;
J. Ellis, A. Ferstl and K.A. Olive, hep--ph/0007113, and 
Phys. Lett. {\bf B481}, 304, (2000), hep--ph/0001005;
V. Mandic, A. Pierce, P. Gondolo and H. Murayama, hep--ph/0008022;
A.B. Lahanas, D.V. Nanopoulos and V.C. Spanos, hep--ph/0009065;
A. Bottino, F. Donato, N. Fornengo and S. Scopel, hep--ph/0010203.

\bibitem{radbreak}
For a review, see e.g. L.E. Ib\'a\~nez and G.G. Ross, in {\it
Perspectives on Higgs physics}, G.L. Kane, editor;
hep--ph/9204201,

\bibitem{dn2}
M. Drees and M.M. Nojiri, Phys. Rev. {\bf D45}, 2482 (1992).

\bibitem{2loop}
J.R. Espinosa and R.--J. Zhang, JHEP {\bf 0003}, 026 (2000),
hep--ph/9912236; 
M. Carena, H.E. Haber, S. Heinemeyer, W. Hollik, C.E.M. Wagner and
G. Weiglein, Nucl. Phys. {\bf B580}, 29 (2000), hep--ph/0001002.

\bibitem{dd}
A. Djouadi and M. Drees, Phys. Lett. {\bf B484}, 183 (2000),
hep--ph/0004205.

\bibitem{bdd} 
C. Boehm, A. Djouadi and M. Drees, Phys. Rev. {\bf D62}, 035012
(2000), hep--ph/9911496. See also J. Ellis, T. Falk, K.A. Olive and
M. Srednicki, Astropart. Phys. {\bf 13}, 181 (2000), hep--ph/9905481.

\bibitem{dy}
M. Drees and A. Yamada, Phys. Rev. {\bf D53}, 1586 (1996),
hep--ph/9508254.

\bibitem{lephiggs}
See e.g. http://lephiggs.web.cern.ch/LEPHIGGS/ .

\bibitem{alnew}
ALEPH collab., R. Barate et al., hep--ex/0011047.

\bibitem{ombound}
See e.g. J.R. Primack, talk at the {\it 4th International Symposium on
Sources and Detection of Dark Matter in the Universe (DM 2000)},
Marina del Rey, California, February 2000, astro--ph/0007187.

\bibitem{dn1}
M.~Drees, and M.~M.~Nojiri, Nucl. Phys. {\bf B369}, 54 (1992).

\bibitem{fenga}
J.L. Feng, K.T. Matchev and F. Wilczek, astro--ph/0008115.

\bibitem{an1}
R. Arnowitt and Pran Nath, Phys. Rev. Lett. {\bf 69}, 728 (1992).

\bibitem{falk}
T. Falk, Phys. Lett. B{\bf 456}, 171 (1999), hep--ph/9902352.

\bibitem{bsg}
S. Bertolini, F. Borzumati, A. Masiero and G. Ridolfi,
Nucl. Phys. {\bf B353}, 591 (1991). 

\bibitem{bsgnlo}
G. Degrassi, P. Gambino and G.F. Giudice, hep--ph/0009337.

\bibitem{string-SG}
A.~Brignole, L.~Iba\~nez and C.~Mu\~noz, Nucl. Phys. {\bf B422}, 125
(1994), Erratum-ibid. {\bf B436}, 747 (1995), hep--ph/9308271;
T.~Kobayashi, D.~Suematsu, K.~Yamada and Y.~Yamagishi,
Phys. Lett.  {\bf B348}, 402 (1995), hep--ph/9408322;
A.~Brignole, L.~Iba\~nez, C.~Mu\~noz and C.~Scheich, 
Z. Phys.  {\bf C74}, 157 (1997), hep--ph/9508258.

\bibitem{sum-rule}
Y.~Kawamura, T.~Kobayashi and J.~Kubo, Phys. Lett. {\bf B405}, 64
(1997), hep-ph/9703320.

\bibitem{dterm}
M. Drees, Phys. Lett. {\bf B181}, 279 (1986);
Y. Kawamura, H. Murayama and M. Yamaguchi, Phys. Rev. {\bf D51}, 1337
(1995), hep--ph/9406245.

\bibitem{dilaton}
V. S. Kaplunovsky and J. Louis, Phys. Lett. {\bf B306}, 269 (1993),
hep--th/9303040.

\bibitem{baertata}
H. Baer, M.A. Diaz, J. Ferrandis and X. Tata, Phys. Rev. {\bf D61}
111701 (2000), hep--ph/9907211.

\bibitem{focus}
J.L. Feng, K.T. Matchev and T. Moroi, Phys. Rev. {\bf D61}, 075005
(2000), hep--ph/9909334.

\bibitem{focus2}
J.L. Feng, K.T. Matchev and F. Wilczek, Phys. Lett. {\bf B482}, 388
(2000), hep--ph/0004043.


\end{thebibliography}
\end{document}